\documentclass[apjl]{emulateapj}
\usepackage{amsfonts,amsmath,graphicx,natbib,apjfonts,subfigure,color,gensymb,longtable}
\usepackage{hyperref}
\usepackage{perpage} 
\MakePerPage{footnote} 

\newcommand{\urlDR}{{\tt https://des.ncsa.illinois.edu/releases/sn}}

\received{November 9, 2018}
\accepted{February 13, 2019}

\def\percentbias{3~mmag}

\shorttitle{\textit{Brout~et~al.}: DES-SN Photometry Pipeline and Y1-Y3 Spectroscopic SNe Ia Light Curve Data Release}
\shortauthors{\textit{Brout~et~al.}: DES-SN Photometry Pipeline and Y1-Y3 Spectroscopic SNe Ia Light Curve Data Release}

\begin{document}
{
\vspace*{-\headsep}\vspace*{\headheight}
\footnotesize \hfill FERMILAB-PUB-18-540-AE\\
\vspace*{-\headsep}\vspace*{\headheight}
\footnotesize \hfill DES-2017-0311
}

\title{First Cosmology Results Using Type Ia Supernovae From the Dark Energy Survey: Photometric Pipeline and Light Curve Data Release}


\def\andname{}

\def\andname{}

\author{
D.~Brout\altaffilmark{1},
M.~Sako\altaffilmark{1},
D.~Scolnic\altaffilmark{2},
R.~Kessler\altaffilmark{3,2},
C.~B.~D'Andrea\altaffilmark{1},
T.~M.~Davis\altaffilmark{4},
S.~R.~Hinton\altaffilmark{4},
A.~G.~Kim\altaffilmark{5},
J.~Lasker\altaffilmark{3,2},
E.~Macaulay\altaffilmark{6},
A.~M\"oller\altaffilmark{7,8},
R.~C.~Nichol\altaffilmark{6},
M.~Smith\altaffilmark{9},
M.~Sullivan\altaffilmark{9},
R.~C.~Wolf\altaffilmark{10},
S.~Allam\altaffilmark{11},
B.~A.~Bassett\altaffilmark{12,13},
P.~Brown\altaffilmark{14},
F.~J.~Castander\altaffilmark{15,16},
M.~Childress\altaffilmark{9},
R.~J.~Foley\altaffilmark{17},
L.~Galbany\altaffilmark{18},
K.~Herner\altaffilmark{11},
E.~Kasai\altaffilmark{19,13},
M.~March\altaffilmark{1},
E.~Morganson\altaffilmark{20},
P.~Nugent\altaffilmark{5},
Y.-C.~Pan\altaffilmark{21,22},
R.~C.~Thomas\altaffilmark{5},
B.~E.~Tucker\altaffilmark{7,8},
W.~Wester\altaffilmark{11},
T.~M.~C.~Abbott\altaffilmark{23},
J.~Annis\altaffilmark{11},
S.~Avila\altaffilmark{6},
E.~Bertin\altaffilmark{24,25},
D.~Brooks\altaffilmark{26},
D.~L.~Burke\altaffilmark{27,28},
A.~Carnero~Rosell\altaffilmark{29,30},
M.~Carrasco~Kind\altaffilmark{31,20},
J.~Carretero\altaffilmark{32},
M.~Crocce\altaffilmark{15,16},
C.~E.~Cunha\altaffilmark{27},
L.~N.~da Costa\altaffilmark{30,33},
C.~Davis\altaffilmark{27},
J.~De~Vicente\altaffilmark{29},
S.~Desai\altaffilmark{34},
H.~T.~Diehl\altaffilmark{11},
P.~Doel\altaffilmark{26},
T.~F.~Eifler\altaffilmark{35,36},
B.~Flaugher\altaffilmark{11},
P.~Fosalba\altaffilmark{15,16},
J.~Frieman\altaffilmark{11,2},
J.~Garc\'ia-Bellido\altaffilmark{37},
E.~Gaztanaga\altaffilmark{15,16},
D.~W.~Gerdes\altaffilmark{38,39},
D.~A.~Goldstein\altaffilmark{40},
D.~Gruen\altaffilmark{27,28},
R.~A.~Gruendl\altaffilmark{31,20},
J.~Gschwend\altaffilmark{30,33},
G.~Gutierrez\altaffilmark{11},
W.~G.~Hartley\altaffilmark{26,41},
D.~L.~Hollowood\altaffilmark{17},
K.~Honscheid\altaffilmark{42,43},
D.~J.~James\altaffilmark{44},
K.~Kuehn\altaffilmark{45},
N.~Kuropatkin\altaffilmark{11},
O.~Lahav\altaffilmark{26},
T.~S.~Li\altaffilmark{11,2},
M.~Lima\altaffilmark{46,30},
J.~L.~Marshall\altaffilmark{14},
P.~Martini\altaffilmark{42,47},
R.~Miquel\altaffilmark{48,32},
B.~Nord\altaffilmark{11},
A.~A.~Plazas\altaffilmark{36},
A.~Roodman\altaffilmark{27,28},
E.~S.~Rykoff\altaffilmark{27,28},
E.~Sanchez\altaffilmark{29},
V.~Scarpine\altaffilmark{11},
R.~Schindler\altaffilmark{28},
M.~Schubnell\altaffilmark{39},
S.~Serrano\altaffilmark{15,16},
I.~Sevilla-Noarbe\altaffilmark{29},
M.~Soares-Santos\altaffilmark{49},
F.~Sobreira\altaffilmark{50,30},
E.~Suchyta\altaffilmark{51},
M.~E.~C.~Swanson\altaffilmark{20},
G.~Tarle\altaffilmark{39},
D.~Thomas\altaffilmark{6},
D.~L.~Tucker\altaffilmark{11},
A.~R.~Walker\altaffilmark{23},
B.~Yanny\altaffilmark{11},
and Y.~Zhang\altaffilmark{11}
\\ \vspace{0.2cm} (DES Collaboration) \\
}

\affil{$^{1}$ Department of Physics and Astronomy, University of Pennsylvania, Philadelphia, PA 19104, USA}
\affil{$^{2}$ Kavli Institute for Cosmological Physics, University of Chicago, Chicago, IL 60637, USA}
\affil{$^{3}$ Department of Astronomy and Astrophysics, University of Chicago, Chicago, IL 60637, USA}
\affil{$^{4}$ School of Mathematics and Physics, University of Queensland,  Brisbane, QLD 4072, Australia}
\affil{$^{5}$ Lawrence Berkeley National Laboratory, 1 Cyclotron Road, Berkeley, CA 94720, USA}
\affil{$^{6}$ Institute of Cosmology and Gravitation, University of Portsmouth, Portsmouth, PO1 3FX, UK}
\affil{$^{7}$ ARC Centre of Excellence for All-sky Astrophysics (CAASTRO)}
\affil{$^{8}$ The Research School of Astronomy and Astrophysics, Australian National University, ACT 2601, Australia}
\affil{$^{9}$ School of Physics and Astronomy, University of Southampton,  Southampton, SO17 1BJ, UK}
\affil{$^{10}$ Graduate School of Education, Stanford University, 160, 450 Serra Mall, Stanford, CA 94305, USA}
\affil{$^{11}$ Fermi National Accelerator Laboratory, P. O. Box 500, Batavia, IL 60510, USA}
\affil{$^{12}$ African Institute for Mathematical Sciences, 6 Melrose Road, Muizenberg, 7945, South Africa}
\affil{$^{13}$ South African Astronomical Observatory, P.O.Box 9, Observatory 7935, South Africa}
\affil{$^{14}$ George P. and Cynthia Woods Mitchell Institute for Fundamental Physics and Astronomy, and Department of Physics and Astronomy, Texas A\&M University, College Station, TX 77843,  USA}
\affil{$^{15}$ Institut d'Estudis Espacials de Catalunya (IEEC), 08034 Barcelona, Spain}
\affil{$^{16}$ Institute of Space Sciences (ICE, CSIC),  Campus UAB, Carrer de Can Magrans, s/n,  08193 Barcelona, Spain}
\affil{$^{17}$ Santa Cruz Institute for Particle Physics, Santa Cruz, CA 95064, USA}
\affil{$^{18}$ PITT PACC, Department of Physics and Astronomy, University of Pittsburgh, Pittsburgh, PA 15260, USA}
\affil{$^{19}$ Department of Physics, University of Namibia, 340 Mandume Ndemufayo Avenue, Pionierspark, Windhoek, Namibia}
\affil{$^{20}$ National Center for Supercomputing Applications, 1205 West Clark St., Urbana, IL 61801, USA}
\affil{$^{21}$ Division of Theoretical Astronomy, National Astronomical Observatory of Japan, 2-21-1 Osawa, Mitaka, Tokyo 181-8588, Japan}
\affil{$^{22}$ Institute of Astronomy and Astrophysics, Academia Sinica, Taipei 10617, Taiwan}
\affil{$^{23}$ Cerro Tololo Inter-American Observatory, National Optical Astronomy Observatory, Casilla 603, La Serena, Chile}
\affil{$^{24}$ CNRS, UMR 7095, Institut d'Astrophysique de Paris, F-75014, Paris, France}
\affil{$^{25}$ Sorbonne Universit\'es, UPMC Univ Paris 06, UMR 7095, Institut d'Astrophysique de Paris, F-75014, Paris, France}
\affil{$^{26}$ Department of Physics \& Astronomy, University College London, Gower Street, London, WC1E 6BT, UK}
\affil{$^{27}$ Kavli Institute for Particle Astrophysics \& Cosmology, P. O. Box 2450, Stanford University, Stanford, CA 94305, USA}
\affil{$^{28}$ SLAC National Accelerator Laboratory, Menlo Park, CA 94025, USA}
\affil{$^{29}$ Centro de Investigaciones Energ\'eticas, Medioambientales y Tecnol\'ogicas (CIEMAT), Madrid, Spain}
\affil{$^{30}$ Laborat\'orio Interinstitucional de e-Astronomia - LIneA, Rua Gal. Jos\'e Cristino 77, Rio de Janeiro, RJ - 20921-400, Brazil}
\affil{$^{31}$ Department of Astronomy, University of Illinois at Urbana-Champaign, 1002 W. Green Street, Urbana, IL 61801, USA}
\affil{$^{32}$ Institut de F\'{\i}sica d'Altes Energies (IFAE), The Barcelona Institute of Science and Technology, Campus UAB, 08193 Bellaterra (Barcelona) Spain}
\affil{$^{33}$ Observat\'orio Nacional, Rua Gal. Jos\'e Cristino 77, Rio de Janeiro, RJ - 20921-400, Brazil}
\affil{$^{34}$ Department of Physics, IIT Hyderabad, Kandi, Telangana 502285, India}
\affil{$^{35}$ Department of Astronomy/Steward Observatory, 933 North Cherry Avenue, Tucson, AZ 85721-0065, USA}
\affil{$^{36}$ Jet Propulsion Laboratory, California Institute of Technology, 4800 Oak Grove Dr., Pasadena, CA 91109, USA}
\affil{$^{37}$ Instituto de Fisica Teorica UAM/CSIC, Universidad Autonoma de Madrid, 28049 Madrid, Spain}
\affil{$^{38}$ Department of Astronomy, University of Michigan, Ann Arbor, MI 48109, USA}
\affil{$^{39}$ Department of Physics, University of Michigan, Ann Arbor, MI 48109, USA}
\affil{$^{40}$ California Institute of Technology, 1200 East California Blvd, MC 249-17, Pasadena, CA 91125, USA}
\affil{$^{41}$ Department of Physics, ETH Zurich, Wolfgang-Pauli-Strasse 16, CH-8093 Zurich, Switzerland}
\affil{$^{42}$ Center for Cosmology and Astro-Particle Physics, The Ohio State University, Columbus, OH 43210, USA}
\affil{$^{43}$ Department of Physics, The Ohio State University, Columbus, OH 43210, USA}
\affil{$^{44}$ Harvard-Smithsonian Center for Astrophysics, Cambridge, MA 02138, USA}
\affil{$^{45}$ Australian Astronomical Optics, Macquarie University, North Ryde, NSW 2113, Australia}
\affil{$^{46}$ Departamento de F\'isica Matem\'atica, Instituto de F\'isica, Universidade de S\~ao Paulo, CP 66318, S\~ao Paulo, SP, 05314-970, Brazil}
\affil{$^{47}$ Department of Astronomy, The Ohio State University, Columbus, OH 43210, USA}
\affil{$^{48}$ Instituci\'o Catalana de Recerca i Estudis Avan\c{c}ats, E-08010 Barcelona, Spain}
\affil{$^{49}$ Brandeis University, Physics Department, 415 South Street, Waltham MA 02453}
\affil{$^{50}$ Instituto de F\'isica Gleb Wataghin, Universidade Estadual de Campinas, 13083-859, Campinas, SP, Brazil}
\affil{$^{51}$ Computer Science and Mathematics Division, Oak Ridge National Laboratory, Oak Ridge, TN 37831}

\begin{abstract}
We present \textit{griz} light curves of 251 Type Ia Supernovae (SNe~Ia) from the first 3 years of the Dark Energy Survey Supernova Program's (DES-SN) spectroscopically classified sample. The photometric pipeline described in this paper produces the calibrated fluxes and associated uncertainties used in the cosmological parameter analysis (\citealt{Brout18-SYS}-SYS, \citealt{KEYPAPER}) by employing a scene modeling approach that simultaneously models a variable transient flux and temporally constant host galaxy. We inject artificial point sources onto DECam images to test the accuracy of our photometric method. Upon comparison of input and measured artificial supernova fluxes, we find flux biases peak at \percentbias. We require corrections to our photometric uncertainties as a function of host galaxy surface brightness at the transient location, similar to that seen by the DES Difference Imaging Pipeline used to discover transients. The public release of the light curves can be found at \urlDR. \\

\end{abstract}

\section{Introduction}
\label{Sec:intro}

\begin{figure*}
\centering
\huge\textbf{SMP Model Visual Representation}\par\medskip
\includegraphics[width=0.75\textwidth]{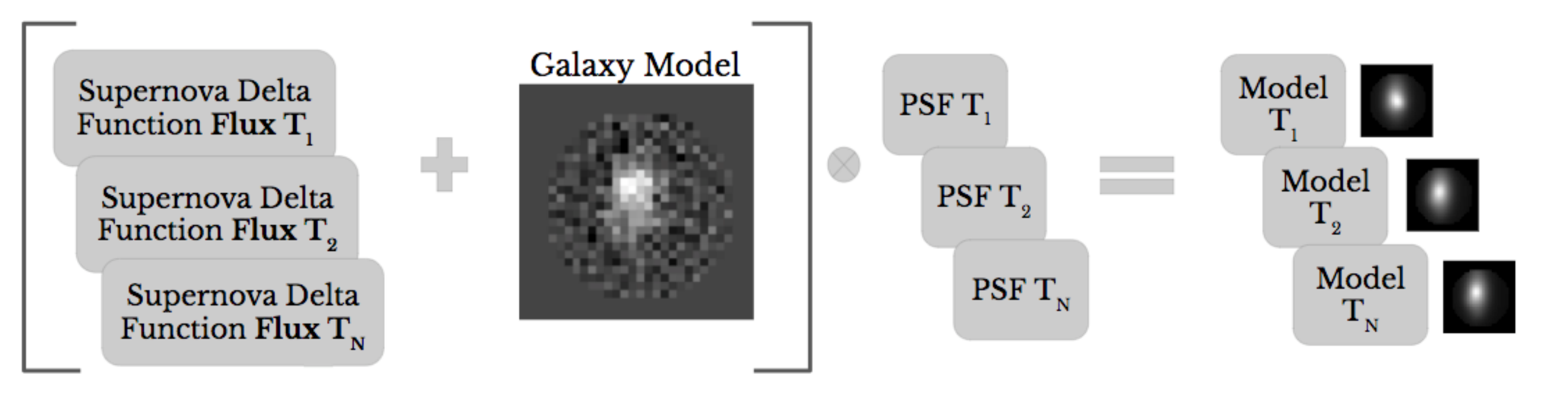}
\caption{Visual representation of the {\tt SMP} process. The model is comprised of a temporally constant galaxy model and a temporally varying SN flux (delta function). Both the SN and galaxy are convolved with the PSF of each image in Fourier space to produce a model which can be compared to data.}
\label{flow}
\end{figure*}

The discovery of the accelerated expansion of the universe (\citealt{riess98}; \citealt{perlmutter99}) using Type Ia Supernovae (SNe~Ia) has motivated the collection of ever-larger SN~Ia samples in order to improve measurements of cosmological distances and test the nature of dark energy. Constraints from SNe~Ia are best measured with a combination of low ($z<0.1$) and higher ($z>0.1$) redshift SNe. The trend in SN surveys over the last three decades has been towards wider and/or deeper rolling surveys where the same images are used to both discover SNe and measure their light curves. The rolling search is conducive to forward modeling photometric methods. So called `Scene Modeling Photometry` (hereafter {\tt SMP}), which  simultaneously models a variable transient flux and temporally constant host galaxy, was first developed by \cite{astier06} and has been implemented for recent SN~Ia cosmology analyses including for the Sloan Digital Sky Survey (SDSS; \citealt{holtzman08}, hereafter H08) and Supernova Legacy Survey (SNLS; \citealt{astier13}, hereafter A13), and as a crosscheck in Pan-STARRS (PS1; \citealt{pantheon}).

The Dark Energy Survey was conducted in two parts; a wide-field galaxy survey (5,000~deg$^2$) and a dedicated transient search in the southern celestial hemisphere covering an area of 27~deg$^2$ (\citealt{des12}, K15: \citealt{kessler15}). The Dark Energy Survey Supernova Program (hereafter DES-SN) has discovered tens of thousands of transients, of which $\sim 3000$ are photometrically classified SNe~Ia covering $0.01<z<1.2$. A subset of $\approx 500$ SNe~Ia from $0.017<z<0.9$ over 5 years has been spectroscopically confirmed. 

In this work we detail and validate our {\tt SMP} pipeline, which forward models SNe and their host galaxies to obtain the DES-SN lightcurves used for cosmological analysis. This paper is part of a series of 9 papers describing the analyses that lead to cosmological constraints from the spectroscopic SNe~Ia observed in the first three years of DES-SN and combined with a low-redshift sample (hereafter DES-SN3YR).
These are: the DES-SN search \& discovery (K15), spectroscopic follow-up (\citealt{D'Andrea18}), calibration (\citealt{Lasker18}), photometry (this work), simulations of our dataset (\citealt{Kessler18}), analysis of Host-SN correlations (Smith et al. in prep.), an inverse distance ladder $H_0$ measurement (\citealt{Macaulay}), the blinded cosmological analysis and systematics validation (\mbox{B18-SYS}: \citealt{Brout18-SYS}-SYS), a Bayesian Hierarchical Method of cosmological parameter fitting (\citealt{Hinton18}), and ultimately the unblinded cosmological parameter constraints (\citealt{KEYPAPER}).

Prior to implementing {\tt SMP}, supernova candidates were discovered and located by the Difference Imaging pipeline, hereafter \texttt{DiffImg} (K15), which uses template images, degrades either the template image or the search image to match the image with worse seeing, and performs an image subtraction to produce catalogs of transient detections. \texttt{DiffImg} then creates candidates from multiple spatially coincident detections, and produces light curves from PSF photometry on the differenced images. \texttt{DiffImg} photometry is used in the real-time analysis of light curves for the spectroscopic follow-up program, and has already been used in several analyses (\citealt{doctor}; \citealt{marcelledark}; \citealt{marcelle}). {\tt SMP} is not used for transient discovery because it would require modeling of all galaxies within the DES-SN footprint, which is not tractable for real-time transient searches. However, because our {\tt SMP} pipeline does not degrade images in the extraction of SN fluxes, it is ideal for use in precision cosmology. The light curves presented here are used in the DES-SN3YR cosmological parameter analysis (\mbox{B18-SYS}) and for obtaining cosmological constraints (\citealt{KEYPAPER}).

We describe our implementation of the scene modeling concept, which is derived from the techniques used by SDSS (H08) and SNLS (A13) and has been developed specifically for DES-SN cosmology. Scene modeling methods have been used extensively in other types of analyses such as crowded-field photometry (\citealt{riessetal16}, \citealt{schlafly18}). In our implementation of {\tt SMP}, the transient flux and host galaxy are modeled simultaneously. The transient flux is allowed to vary over time and the host galaxy flux is fixed across all observations. 

In order to evaluate the results of scene modeling photometry, A13 moved nearby stars on their images to locations near host galaxies and treated them as fake SNe but did not measure light curves. We have developed a unique approach in which we generated 100,000 artificial SN light curves that are inserted as point sources onto DECam images (hereafter `fakes'). Injection of artificial point sources is one component of a multi-faceted plan to use fake SNe to trace biases throughout the DES-SN cosmological parameter analysis. Here, they are used to check for flux biases introduced by the photometric pipeline and to determine corrections for {\tt SMP} flux uncertainties. \mbox{B18-SYS} use fakes to characterize the output of \texttt{DiffImg} and {\tt SMP}, which is needed for catalog-level simulations that are used to predict distance biases. B18-SYS also present a full cosmological analysis of 10,000 fake SNe that have been ``discovered'' by the search pipeline, processed by the {\tt SMP} pipeline, and processed through our cosmological analysis pipeline in the same manner as the real dataset.

One outstanding problem in SN photometry that was dealt with in previous surveys (e.g., R14: \citealt{rest14}, J17: \citealt{jones17}) is the underestimation of SN flux uncertainties when SNe are located near high local host galaxy brightness. R14 and J17 characterize the size of this effect by performing photometry at the location of the SNe when the SN flux is known to be zero. Here, we describe how we use our extensive pipeline of fakes to assess the size of this effect for our analysis and model it precisely in catalog level simulations of our dataset.

The outline of this paper is as follows. We discuss our dataset, the preparation, and internal calibration of DES images in Section \ref{Sec:data}. Our scene-modeling method is explained in Section \ref{Sec:method}. In Section \ref{fakes} we show the results of our validation on fakes. In Section \ref{data} we apply our pipeline to the DES-SN 3 year spectroscopic sample and present the light curves used for our cosmological parameter analysis; the publicly released light curve data can be found online\footnote{DES-SN~Spectroscopic~Sample~Y1-Y3~{\tt SMP}~Photometry~Release: \urlDR}. In Section \ref{psfcrosscheck} we crosscheck the PSF model because it is not tested in our fakes analysis. In Section \ref{discussion} we discuss improvements to {\tt SMP} and we compare to \texttt{DiffImg} and in Section \ref{conclusion} we give our conclusions.

\section{Dataset and Image Pre-processing}
\label{Sec:data}

\subsection{The 3 Year Spectroscopic Sample}

The DES-SN performed a deep, time-domain survey in four optical bands $(griz)$ with an average cadence of 7 days per filter covering $\sim 27$~deg$^2$ over 5 annual campaigns from 2013 to 2018 using the Dark Energy Camera (DECam: \citealt{decam}). DECam exposure processing (\citealt{morganson}), \texttt{DiffImg}, and automated artifact rejection (\citealt{autoscan}) were run on a nightly basis.

DES-SN observed in 8 ``shallow'' and in 2 ``deep'' fields, with the shallow and deep fields having typical nightly point-source depths of 23.5 and 24.5~mag, respectively.  Multiple exposures are taken each night with 3, 3, 5, and 11 (1, 1, 1, and 2) exposures taken in $griz$ for the deep (shallow) fields (See \citealt{D'Andrea18}).
Images used in this analysis were taken during the first three years of DES-SN, from Sept. 2013 to Feb. 2016, in which we discovered roughly $\sim$12,000 transients. Among these transients, $\sim$3,000 were identified as likely SNe~Ia based on their light curves and 251 were spectroscopically confirmed (\citealt{D'Andrea18}).

\subsection{Image Processing}
\subsubsection{FirstCut}

The DECam images used by the {\tt SMP} pipeline are first pre-processed as part of the nightly single-epoch processing. This pre-processing stage, denoted FirstCut (\citealt{morganson}), accounts for crosstalk correction, bias subtraction, bad-pixel masking (masking known problematic pixels in the camera), and flat fielding. It also makes corrections to image fluxes for CCD nonlinearity (\citealt{des_bernstein_astrometry_17}) and the brighter-fatter effect (A13, \citealt{antilogus}, and \citealt{gruen15}), and it masks cosmic rays and satellite trails. 

A sky level has been fit and subtracted using the principle component analysis pipeline developed by \cite{bernsteinInstrumentResponse}. This procedure decomposes the image under the assumption that it is the sum of the astrophysical sources of interest, a zero-mean noise component, and a background component that is a linear function of a small number of sky templates.

\subsubsection{Additional Image Preparation}
After FirstCut, we perform additional image preparation. While we do not use \texttt{DiffImg} photometry, we use a number of the same modules as summarized below and described in detail in K15.
For each exposure and CCD we perform the following steps: 
i) compute an astrometric solution from a joint fit to a template image, resulting in improved relative astrometry between the different epochs,
ii) determine a position-dependent PSF following the K15 options instead of those from FirstCut, and 
iii) overlay the same fakes that were overlaid during the search. Additionally, we use a DES-derived stellar catalog (described in Section 3.2.1 of K15.) instead of an external catalog such as USNO-B (\citealt{monet03}).

\subsection{Star Catalog}
\label{Sec:starcatalog}
Calibrated tertiary standard star magnitudes from \cite{burke} are used for the DES-SN internal calibration of each DES-SN image. Approximately 50 tertiary standard stars lie within each DECam CCD image. \cite{burke} have determined \textit{grizY} magnitudes in the AB system of these standard stars using the ``Forward Global Calibration Method'' (FGCM). The FGCM ``forward'' computes the fraction of photons observed for each star over repeated exposures by utilizing measurement of the instrument transmission function, precipitable water vapor, observing conditions, and a model of the stellar source. In addition, using the passband transmission (instrument + atmosphere) vs. wavelength and the spectral energy distribution (SED) of the source, corrections are applied to the stellar catalog fluxes (as well as to the final SN fluxes). These SED-dependent ``chromatic corrections'' account for 
differences between SED and the mean stellar SED, and between
atmospheric transmission of each exposure and the mean atmospheric transmission. This correction extends the FGCM calibration precision to be valid over a wide color range ($-1$ $\lesssim$ $g-i$ $\lesssim$ 3).  We refer the reader to \cite{chromatic} for more detail. The implications on cosmological measurements due to these corrections are discussed in \cite{Lasker18}.

\section{Method}
\label{Sec:method}

The {\tt SMP} method utilizes a set of calibrated DECam image stamps centered at the location of a SN to constrain a model for a temporally varying SN and a temporally constant host galaxy (Figure \ref{flow}). Here we outline the steps required to build and fit the {\tt SMP} model.

\subsection{Stellar Photometry}
\label{stellarphotometry}
We use PSF-fitted photometry of the tertiary standard stars to determine the zero-point of each image. As discussed above, the sky background in the FirstCut images was subtracted using PCA over the entire exposure. However, at the specific locations of transient objects we check for residual nonzero sky background. Residual sky often occurs when the moon is bright, causing large sky gradients that are not captured with PCA. We apply a second method of local sky background and sky uncertainty estimation using concentric apertures of 40 and 60 pixels following \cite{pythonphot} and the resulting sky and uncertainty are calculated in the same manner for each tertiary standard star as well as for the SN.

Biases are induced in PSF-fitted flux measurements when the astrometric solution of a source is incorrect or is uncertain (\citealt{rest14}). These biases are smaller for stars than for SNe because the stars have higher S/N and their positions are better constrained. When computing photometric magnitudes, in the limit of high S/N and a correct PSF model, there is no astrometrically-induced flux bias if the astrometric solution and uncertainty are the same for both the stars and the SN itself. The bias in the zeropoint and the bias in the SN flux will cancel.
Here, we discuss the expected photometric biases in the real SNe dataset; in the fake dataset this is more subtle and is discussed in Sections \ref{biases}. 

There are fundamental differences between stars and the SNe that must be accounted for. The stars may have measurable proper motion while the SNe do not. Additionally, the centroids of SNe have larger uncertainty because there are fewer epochs to constrain the position and the S/N is lower. Therefore, in modeling the SNe, we fix the location of the SN in R.A. and Dec. across all images (Section \ref{Sec:method}). While the SN position is fixed (``fixed-position photometry''), we determine the position of the stars for each image in order to account for stellar proper motions (``variable-position''). Proper motions of the standard stars, which are estimated by linear fits to the positions over 3 years of observations, have an RMS of $\sim 10$~mas per year. 

In order to be consistent in the application of the stellar position in the photometry, \cite{rest14} and \cite{pantheon} run fixed-position photometry on both the stars and SNe. In our pipeline we apply fixed-position photometry to the SN but we apply variable-position photometry on the stars, and this inconsistency causes a small $1-2$~mmag bias towards fainter SN flux measurements but has the benefit of accounting for stellar proper motions.  
These small biases are not corrected for, but rather are incorporated into the systematic uncertainty budget as they are sub-dominant to the total calibration uncertainties of the systematic error budget described in B18-SYS.

\begin{figure}[!h]
\includegraphics[width=0.5\textwidth]{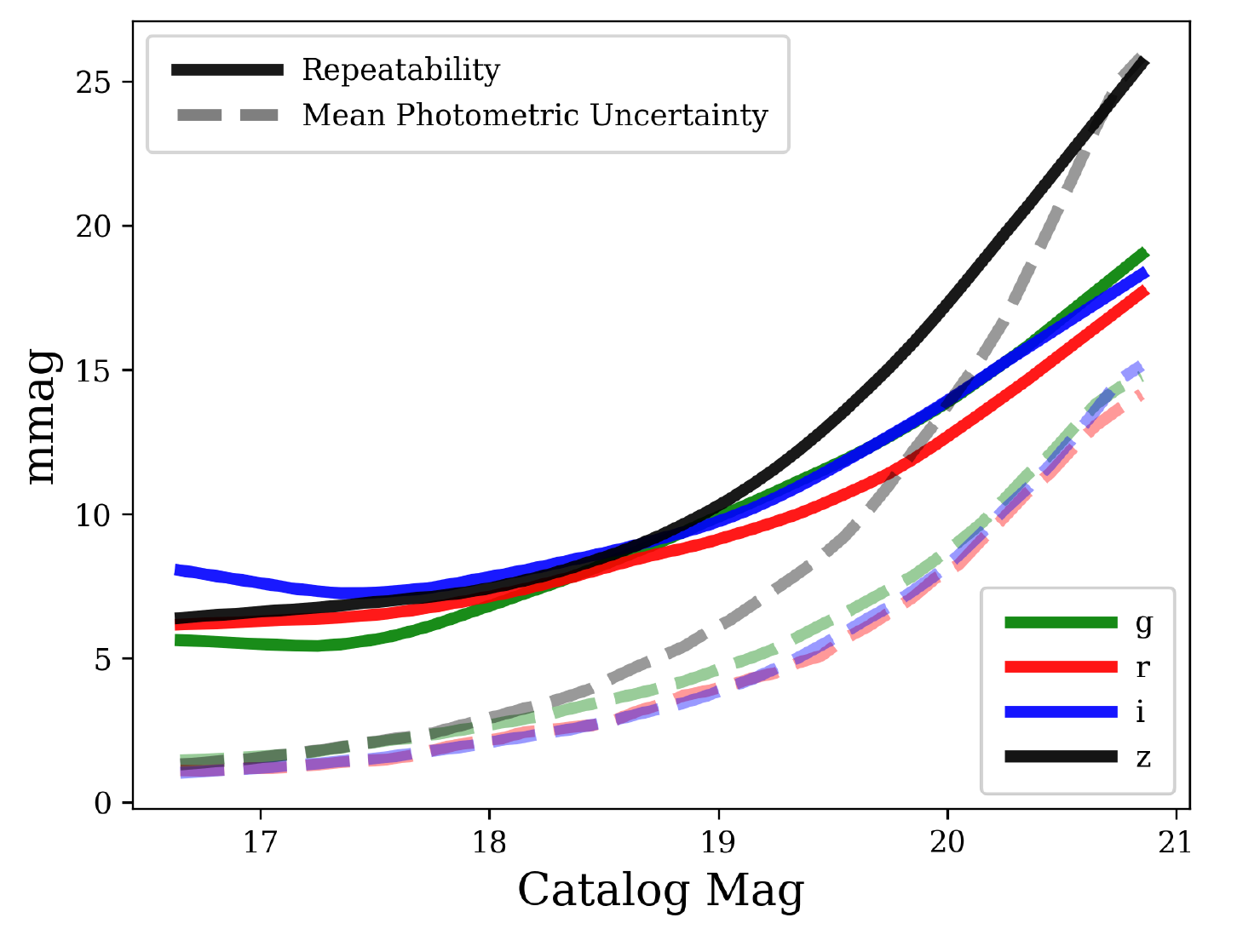}
\caption{Solid lines designate 1$\sigma$ scatter in the recovered stellar magnitude (repeatability) as a function of stellar catalog magnitude for each DECam band. Dotted lines designate the mean photometric uncertainties. There is a floor in the photometric repeatability of $\sim$6~mmag.\\}
\label{fig:floor}
\end{figure}

Millions of tertiary standard star measurements are taken over the course of DES-SN. Following A13, the uncertainty used in the stellar photometry fits does not include source Poisson noise. The $1\sigma$ scatter in the recovered stellar magnitudes (hereafter `repeatability') is plotted in Figure \ref{fig:floor}. For the brightest stars ($<17$~mag), the photometric uncertainties after including Poisson noise analytically are $1$~mmag, but the observed measurement scatter is $>5$~mmag (Figure \ref{fig:floor}) in each band. 
This floor, after subtracting out the mean photometric uncertainty, is added in quadrature to all flux uncertainties.

\begin{figure}[!h]
\centering
\includegraphics[width=0.48\textwidth]{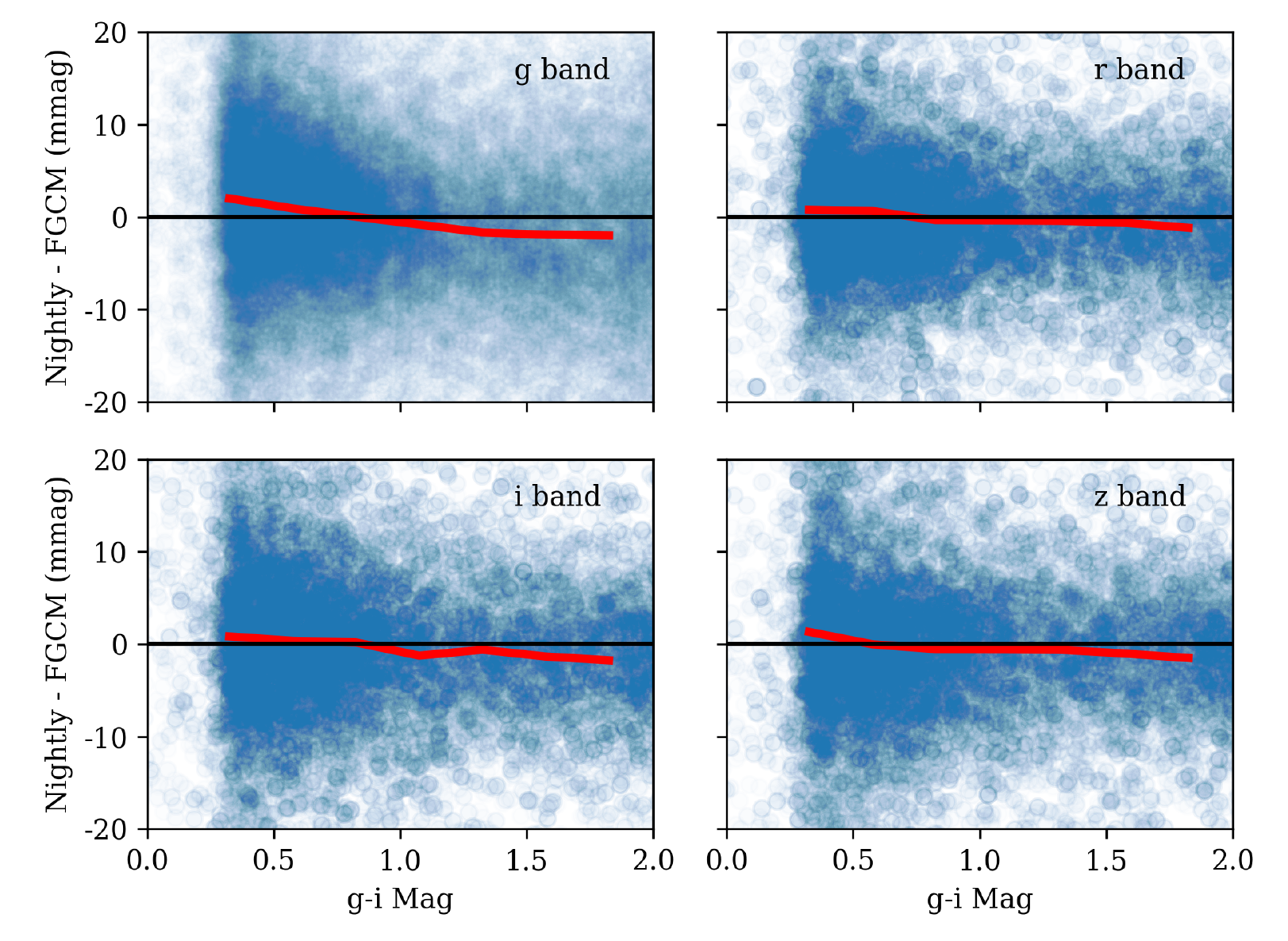}
\caption{Nightly (per exposure) tertiary standard star magnitudes compared to the FGCM pipeline catalog magnitudes as a function of the FGCM catalog $g-i$ color. The color binned mean of the magnitude residuals is shown in red.}
\label{colortrends}
\end{figure}

In order to demonstrate the size of the chromatic corrections applied to the tertiary standards in the SN fields, we compare the un-corrected individual exposure (nightly) stellar photometry with the FGCM chromatically corrected stellar catalog magnitudes (Figure \ref{colortrends}). Differences are up to 4~mmag over the color range of the tertiary standards ($0.25<g-i<2$ mag).

\subsection{Image Model Fitting}
\label{modelfit}

\begin{figure*}
\centering
\includegraphics[width=0.97\textwidth]{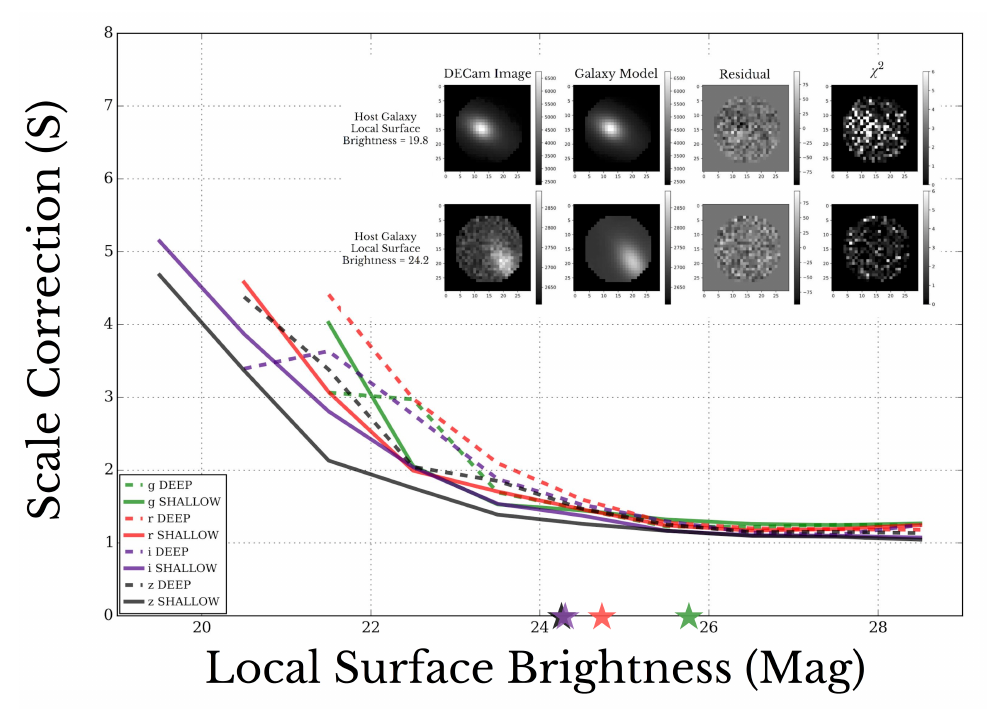}
\caption{Scale Correction (S) = RMS($\Delta F/\sigma_{\rm stat}$) as a function of $m_{SB}$, for 10,000 Fake SNe~Ia processed by {\tt SMP}. The stars on the x-axis denote the mean local surface brightness in the DES subset for each band. Inset: Examples of high and low $m_{SB}$ galaxies and {\tt SMP} best fit models, data $-$ model, and $\chi^2$. }
\vspace{20pt}
\label{fakescale}
\end{figure*}

As in H08 and A13, {\tt SMP} uses a time series of image stamps from the data located at the position of the SN. We assume that the DECam pixel fluxes can be modeled from a temporally varying SN flux and a temporally constant galaxy model that is modeled as a grid of pixels. In order to facilitate model comparisons to all images simultaneously, all data images are scaled to a common zeropoint of 31.00~mag.\footnote{This ZP of 31.00 is for internal {\tt SMP} computations only; the ZP in the public data files is 27.5.}
Following H08 and A13, the model is re-sampled to compare with the dataset and the data are never re-sampled to avoid correlated noise. A visual representation of the model is shown in Fig. \ref{flow}. The ``Model'' images shown on the right hand side of Fig.\ \ref{flow} are compared to data, and to constrain our model we minimize the following:

\begin{equation} \label{chisq}\\
\chi^2 = \sum_{ij,n} \frac{(S_{ij,n}-D_{ij,n})^2}{\sigma_{sky_n}^2},\\
\end{equation}
for each pixel labeled with indices $i$ and $j$, and exposure $n$. $S_{ij,n}$ are the modeled pixel fluxes and $D_{ij,n}$ are the data pixel fluxes. Equation \ref{chisq} is weighted by the pre-computed variance in the sky counts ($\sigma_{sky_n}^2$) as motivated by A13 to preserve statistical optimality for faint sources and avoid potential biases due to inaccuracy of the PSF model. However, because the denominator of Eq.~\ref{chisq} does not include all sources of noise, we modify the photometric uncertainties output by {\tt SMP} using both the analytical expectations of source and galaxy noise (Section \ref{uncertainties}), and we correct our uncertainties using results on fake SNe (Section \ref{fakescale}).

For our model $S_{ij,n}$, we define a temporally varying SN flux for each exposure $n$ ($F_n$) and a temporally constant grid of fluxes ($g_{ij}$) of size $N\times N$ ($N=30$). The SN and host galaxy fluxes per pixel are defined as follows:

\begin{equation}\label{fsn}
FSN_{ij,n} = \pmb{F_n} \sum_{k_i k_j}\tilde{PSF}_{k_i,k_j,n}~ e^{2{\rm i}\pi k_i( \pmb{\bar{SN}_i}-i_0)/N}~e^{2{\rm i}\pi k_j( \pmb{\bar{SN}_j}-j_0)/N},
\end{equation}

\begin{equation}\label{fgal}
FGAL_{ij,n} = \sum_{k_i k_j}\tilde{PSF}_{k_i,k_j,n}~ \pmb{\tilde{g}_{k_i, k_j}}~ e^{-2{\rm i}\pi k_i / N}~e^{-2{\rm i}\pi k_j / N},
\end{equation}\\

and the model image $S_{ij,n}$ is defined as
\begin{equation}
S_{ij,n} =   FSN_{ij,n} + FGAL_{ij,n}, \\
\end{equation}\\
where $\tilde{PSF}_{ij,n}$ is the Fourier transform of the PSF evaluated at the location of the SN.  We vary the SN sky position in Fourier space, where the SN point source is represented by a plane wave at $\bar{SN}_{i}$, $\bar{SN}_{j}$ in pixel coordinates relative to the center of the galaxy model ($i_0$ and $j_0$) which is defined to be the \texttt{DiffImg} SN position. This formalism allows us to model the SN position at sub-pixel locations in Fourier space and to evaluate the likelihood in real space. The floated parameters in our fits are designated in bold font in Equations \ref{fsn} \& \ref{fgal}; these parameters are $F_n$, $g_{ij}$, $\bar{SN}_{i}$, and $\bar{SN}_{j}$.

We adopt a galaxy model on a grid of pixels with the same 0.27\arcsec pixel scale as the DECam images. The reference center of each data stamp is the position of the SN as determined by \texttt{DiffImg}. This position is an average of all epochs for which there was a \texttt{DiffImg} detection. The reference center is at a sub-pixel location, so as to facilitate comparison of our model with the data, we shift the galaxy model and the SN model for each exposure by the difference of the center image pixel and the reference center.

In order to avoid degeneracies between the galaxy model and the SN flux, we fix the model SN flux at zero for epochs outside the observer frame range $\Delta MJD_{\rm peak} > 300$ days or $\Delta MJD_{\rm peak} < -60$ days where $MJD_{\rm peak}$ is the derived date of peak flux from an initial light curve fit of \texttt{DiffImg} photometry and $\Delta MJD_{\rm peak} =  MJD_{\rm exposure} - MJD_{\rm peak}$. We find that any residual SN flux beyond 300 days contributes to negligible biases in photometry ($< 0.01\%$).

\subsection{Implementation}
\label{computing}

We utilize a Markov Chain Monte Carlo Metropolis Hastings algorithm (\citealt{metropolis1953}, \citealt{hastings1970}) to sample the likelihood and we assume flat priors on each of our model parameters with the exception of the SN R.A. and Dec. for which we assume a top-hat prior with radius 2 pixels that is centered at the location of the \texttt{DiffImg} fit sky position. For our model image stamps, we adopt a radius of 13 pixels (3.5 arcsec) around $i_0,j_0$, inside of which we compute $\chi^2$ from Eq.~\ref{chisq} using only pixels that fall entirely within the pre-defined radius. For each filter, we have a total of $\sim$500 galaxy model parameters and anywhere from 25 to 500 SN flux parameters; one for SN flux in each exposure that falls within our defined MJD range over which we fit SN fluxes. For our sampling algorithm, we do not employ more complicated algorithms such as \texttt{emcee} because the computation requirements of our likelihood and the number of parameters make running the required 2N walkers intractable. Instead, during the first 100,000 steps we optimize our steps in each parameter to achieve between 25\% and 75\% acceptance rate. We employ a Geweke Diagnostic (\citealt{geweke}) test to ensure that our chains for the SN fluxes have sufficiently sampled the posterior space. Our chains can run up to 2,000,000 steps. The galaxy model, which is represented as a grid of delta functions in Fourier space, has power on all scales which can lead to poor convergence. For this reason we do not explicitly check for convergence of $g_{ij}$, but rather we ensure convergence of the $FGAL_{ij,n}$ pixels in a 1\arcsec aperture centered at the location of the SN.

The {\tt SMP} fits are performed separately in each band. While there could be added benefit in measuring the SN position by fitting all bands simultaneously, atmospheric refraction causes the position of the SN to be color dependent, which is not accounted for in this work. A total of 41,004 jobs were run independently in order to produce $griz$ light curves for the 251 SNe in the spectroscopic sample and 10,000 fakes. Each job utilized a single FNAL processor and could take anywhere from 5 to 48 hours to fit, with the latter occurring for deep-field  $z$-band fits with up to 750 exposures. The vast majority of the computation time is in the convolution of the galaxy model with the PSF for each exposure. To improve fitting speed, the PSFs were stored in Fourier space and the galaxy model ($g_{ij}$) is transformed to Fourier space and subsequently convolved with the PSF requiring only $n+1$ Fourier transforms. After fitting, we evaluate the best fit $F_{n}$ for each exposure $n$ by taking the mean of the MCMC chain. The error on $F_{n}$ is the standard deviation of the MCMC chain. For observation sequences with multiple back-to-back exposures, 
we report the weighted average flux and uncertainty among the
individual exposures.

\begin{figure*}
\centering
\includegraphics[width=0.8\textwidth]{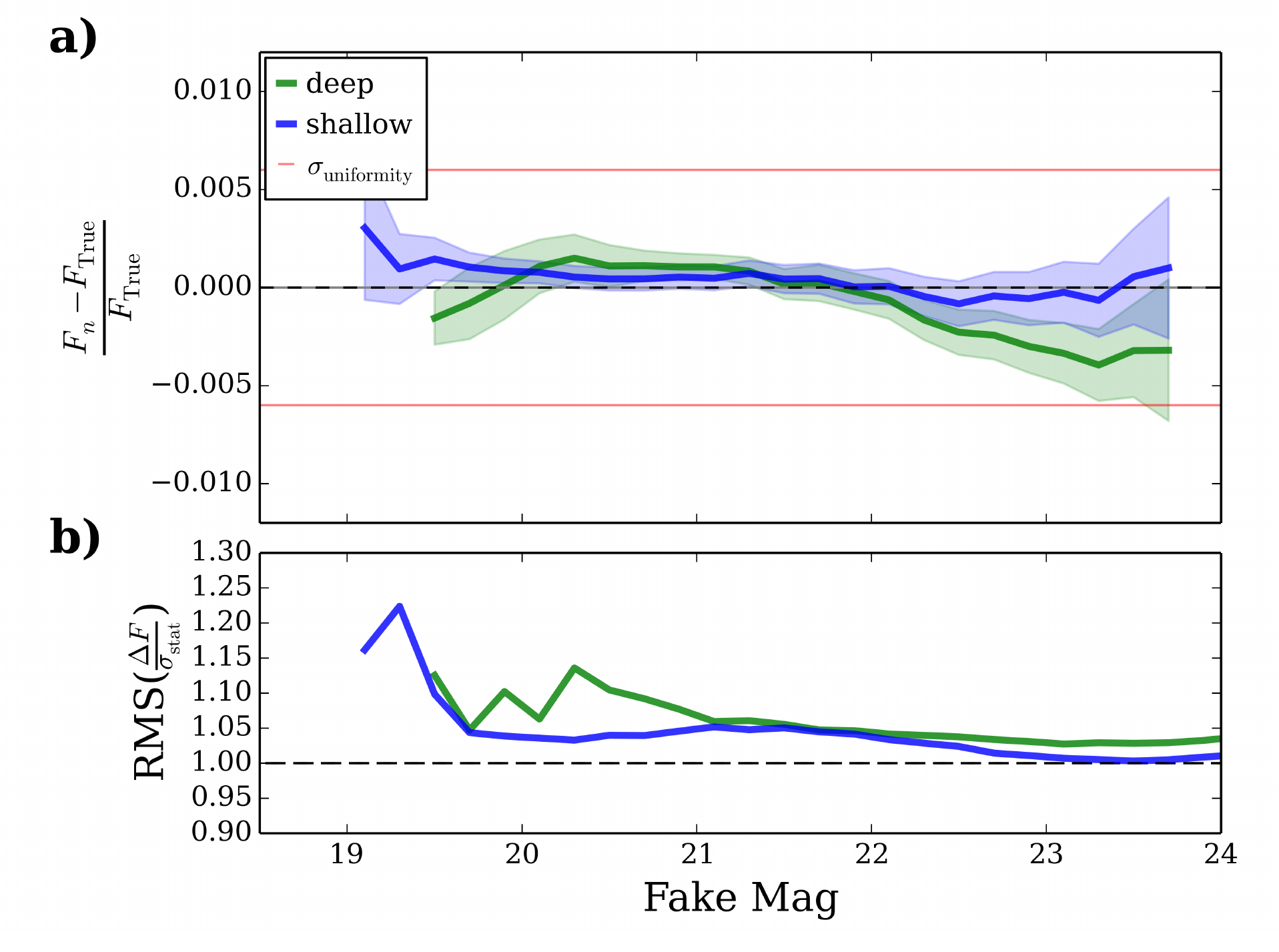}
\caption{a) Fractional flux residuals as a function of fake SN magnitude. All host galaxy local surface brightnesses are included. Comparison with the uncertainty in calibration non-uniformity from \cite{burke} ($\sigma_{\rm uniformity}=0.006$~mag) is shown. The shaded regions designate the 1$\sigma$ errors on the mean.  b) RMS of the pull-distribution as a function of fake SN magnitude. }
\label{fakeresid}
\end{figure*}

\subsection{Uncertainties}
\label{uncertainties}

Here we describe the treatment of the statistical uncertainties within {\tt SMP} to which an additional empirically observed dependence on host galaxy surface brightness is included in Section \ref{fudges}. There has been debate about the proper way to include Poisson noise of the host galaxy and source in the photometry fits (H08 and A13). H08 weight their fits according to expected photon statistics, which includes the Poisson noise of the host galaxy. A13 exclude the noise contribution of the host galaxy and source in the fitting process. We have chosen the latter method (shown in Eq.~\ref{chisq}) and correct our output uncertainties using expected photon statistics after the fitting process following: 

\begin{equation}
\label{sigstat}\\
\sigma_{\rm stat}^2 = \sigma_{\rm {\tt SMP}fit}^2 + \sigma_{\rm source}^2 + \sigma_{\rm hostgal}^2 ,
 \\ \\
\end{equation}
where $\sigma_{\rm {\tt SMP}fit}$ is the uncertainty derived from the {\tt SMP} Monte Carlo chains which were computed using only the sky uncertainty, $\sigma_{\rm source}$ is the Poisson noise of the SN, and $\sigma_{\rm hostgal}$ is the host galaxy Poisson noise. The host galaxy photon variance on exposure $n$ is approximated by

\begin{equation}
\\\sigma^2_{{\rm hostgal},n} = \frac{\sum_{ij} fgal_{ij,n} \times PSF_{ij,n}^2}{\sum_{ij}PSF_{ij,n}^2} \times NEA ,\\
\end{equation}
where $fgal_{ij,n}$ is $FGAL_{ij,n}$ expressed in photoelectrons following: 

\begin{equation}
fgal_{ij,n} = FGAL_{ij,n} \times 10^{(ZP_{n}-31)/2.5} \times {\rm Gain}_n ,\\
\end{equation}
and the noise equivalent area is $NEA \equiv 1/\sum_{ij}PSF_{ij,n}^2$. Equation \ref{sigstat} corresponds to our analytic expectation of the photometric uncertainties. Finally, we report the weighted average uncertainty among the
individual back-to-back exposures. Below we test the accuracy of our photometric extraction and correct $\sigma_{\rm stat}$ for underestimation of the measurement noise.

\section{Corrections and Tests on Fake Supernovae}
\label{fakes}

Fake SN Ia light curves are inserted onto DECam images at locations of real galaxies. Here we analyze a set of 10,000 fakes that were discovered by \texttt{DiffImg} and processed by {\tt SMP}. We optimize our pipeline for minimal photometric outliers, check for biases in our photometric method, and apply corrections to our photometric uncertainties.

\subsection{Fake Supernovae}

The insertion of fake SNe at the image level and the subsequent analysis of their measured fluxes is an important test of the photometric pipeline. It allows us to quantify measurement biases, compare {\tt SMP} uncertainties to the measured minus true flux differences and determine uncertainty corrections, and optimize {\tt SMP} cuts to reject flux outliers. We simulate a sample of SN~Ia light curves and insert light curve fluxes onto DES-SN images using the measured PSF. Because we insert an entire sample of SN~Ia light curves, we are able to characterize biases in photometry as well as the propagation of these photometry biases to biases in measured distances.
A13 moved nearby stars in their images to locations near host galaxies and treated them as fake transients, which preserves the true PSF for each star, but it is difficult to trace photometry biases to distance biases given that they have limited statistics of fake stars and do not model a sample of fake SNe light curve magnitudes. Additionally, A13 did not account for a position-dependent PSF when moving stars, whereas the method described here does.

Fake SN light curve fluxes are generated using the SuperNova ANAlysis software package (\texttt{SNANA}: \citealt{snana}) in a $\Lambda$CDM cosmology ($\Omega_M$=0.3). Light curve fluxes are overlaid as PSF sources onto the DECam images and processed with the \texttt{DiffImg} pipeline. A detailed description of the simulation used for the fakes can be found in Section 2 of \cite{Kessler18}, but here we provide a brief summary. The fake SNe span a wide magnitude range (from $19^{th}$~mag to well below the detection limit) and redshift range ($0.1 < z < 1.2$). K15 overlay fluxes onto the CCD image near real galaxies with SN locations chosen with a probability proportional to the host surface brightness density. The SN flux is distributed over nearby pixels using the PSF determined with \texttt{PSFEx}, and the flux in each pixel is varied by random Poisson noise.  Since we use a scaling of the modeled PSF to insert the fake transient, rather than the real PSF (i.e. moving real stars in the image), we separately check for potential PSF modeling errors that are not included as a part of the analysis of the fakes.

K15 inserted 100,000 fake SN light curves into the first 3 years of DES-SN images. These fakes were used to monitor image quality and $\sim$40,000 fake SNe~Ia were ``discovered'' by \texttt{DiffImg}. However because {\tt SMP} is computationally expensive, for this first DES-cosmology analysis, only on a subset of 10,000 fake SN light curves were processed by {\tt SMP}.

\subsection{Outlier Rejection}
In order to reduce the number of photometric outliers, exposure quality requirements (cuts) were optimized on the sample of fake SNe. We denote the fraction of 5$\sigma$ flux outliers ($\eta_{5\sigma}$) when comparing the {\tt SMP} fit flux ($F_n$) to the true fake flux ($F_{\rm True}$). We remove exposures with poor data-model agreement ($\chi^2$/ndof$>1.2$) and with poor seeing conditions (PSF$_{\rm FWHM} >$ 2.75 arcsec). To make additional improvements we also place conservative cuts based on zeropoint and sky level to remove the poorest quality images. These cuts retain 94\% of all exposures and reduce $\eta_{5\sigma}$ from $6\times10^{-4}$  to $2\times10^{-4}$.

\subsection{Photometry Biases}
\label{biases}

Comparing the input photometry to the recovered photometry ($\Delta F = F_n - F_{\rm True}$), we measure photometric biases $< 0.5$\% over 19th to 24th magnitude. As shown in panel a) of Figure \ref{fakeresid}, there is a slight bias in the deep fields for $\Delta F/F_{\rm True}$ of $-0.3$\% at faint magnitudes, which is included in the systematic error budget of B18-SYS. 

There are three key differences between the analysis of the DES-SN dataset and that of the fake SNe. First, the astrometric solution used to insert fakes (K15) is the same solution that is used to model the fakes within {\tt SMP}. Astrometric uncertainty is not simulated in the fake point sources.  Second, K15 use zeropoints that were fit using aperture photometry to insert fake fluxes onto images, while {\tt SMP} uses PSF fitting. In order to assess the accuracy of {\tt SMP}, we correct for the zeropoint difference between the K15 and {\tt SMP}. Thus, our results presented here are insensitive to incorrect modeling of the zeropoint. B18-SYS discuss an independent method for validating the zeropoint and internal calibration uncertainties. Third, the analysis of the fakes uses the same PSF model that was used to insert the fakes. Inaccuracies of the PSF model are not simulated in the fakes, and thus in Section \ref{psfcrosscheck} we perform a crosscheck of our PSF model.

If the {\tt SMP} flux uncertainties are accurate, then RMS$(\Delta F/\sigma_{\rm stat})=1$. However, we observe that the RMS of the fakes is slightly above unity as shown in panel b) of Figure \ref{fakeresid}. To characterize the excess scatter, we examine the dependence of the RMS on the local host galaxy local surface brightness ($m_{SB}$).

\subsection{Host Galaxy Surface Brightness Dependence}
\label{fudges}
We find that there is an underestimation of photometric uncertainties for SNe located in galaxies with high local surface brightness, as was seen previously in \texttt{DiffImg} (K15). A scale correction ($S$) is computed from the fakes as shown in Figure \ref{fakescale} that is required to bring RMS of recovered fake fluxes as a function of $m_{SB}$ to unity. This dependence (hereafter the Host SB dependence) has been seen in the past (K15, \citealt{pantheon}).  The source of the Host SB dependence is unclear since we include host galaxy Poisson noise in our {\tt SMP} uncertainty calculation (see Sec. \ref{uncertainties}). In {\tt SMP}, we find no significant bias in $\Delta F/\sigma_{\rm stat}$ as a function of $m_{SB}$.
 
The inset of Figure \ref{fakescale} shows the results of {\tt SMP} run on two example host galaxies, one bright and one faint. For the bright host galaxy, visibly poorer $\chi^2$ distributions are seen across the image stamp and structure can be seen in the residual stamp.

To account for the increased scatter as a function of host galaxy surface brightness, K15 scaled their output SN flux uncertainties. In {\tt SMP} we apply the same method of scaling our SN flux uncertainties with multiplicative corrections ($S$). The {\tt SMP} light curve photometric uncertainties ($\sigma_{F}$) are given by

\begin{equation}\\\sigma_{F} = \sigma_{\rm stat} \times S
 \\ \\
\end{equation} where $\sigma_{\rm stat}$ was defined as the co-added measurement uncertainty and $S$ is the function of $m_{SB}$, bandpass, and field shown in Figure \ref{fakescale}.

\begin{figure*}[!h]
\begin{centering}

\includegraphics[page=1,trim={0 18cm 0 0},clip,width=0.49\textwidth]{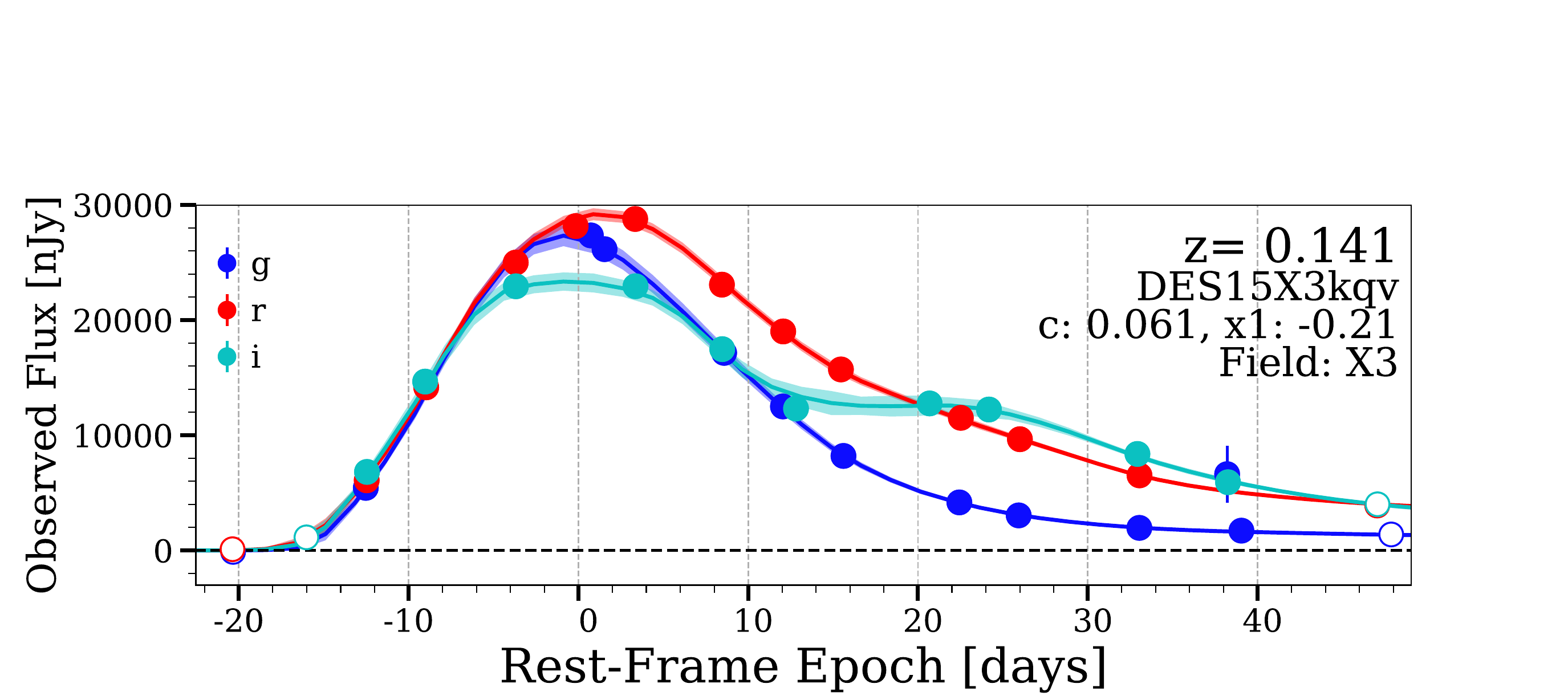}
\includegraphics[page=2,trim={0 18cm 0 0},clip,width=0.49\textwidth]{representativelc6_withc_and_x1_shortened.pdf}
\includegraphics[page=3,trim={0 18cm 0 0},clip,width=0.49\textwidth]{representativelc6_withc_and_x1_shortened.pdf}
\includegraphics[page=4,trim={0 18cm 0 0},clip,width=0.49\textwidth]{representativelc6_withc_and_x1_shortened.pdf}
\includegraphics[page=5,trim={0 18cm 0 0},clip,width=0.49\textwidth]{representativelc6_withc_and_x1_shortened.pdf}
\includegraphics[page=6,trim={0 18cm 0 0},clip,width=0.49\textwidth]{representativelc6_withc_and_x1_shortened.pdf}
\includegraphics[page=7,trim={0 18cm 0 0},clip,width=0.49\textwidth]{representativelc6_withc_and_x1_shortened.pdf}
\includegraphics[page=8,trim={0 18cm 0 0},clip,width=0.49\textwidth]{representativelc6_withc_and_x1_shortened.pdf}
\includegraphics[page=9,trim={0 18cm 0 0},clip,width=0.49\textwidth]{representativelc6_withc_and_x1_shortened.pdf}
\includegraphics[page=10,trim={0 18cm 0 0},clip,width=0.49\textwidth]{representativelc6_withc_and_x1_shortened.pdf}
\par\medskip\medskip
\caption{Representative light curves of DES SNe from the DES-SN3YR sample with photometric data provided by {\tt SMP} and fits to the light curve data provided by SALT2 simply intended to guide the reader's eye. SNe with C3 or X3 in the name are found in deep fields, the remaining SNe are found in the shallow fields. The fields are described in detail Section 2.1 of B18-SYS. \\}

\vspace*{+0.5in} 
\label{representativelc}
\end{centering}
\end{figure*}

\begin{figure}
\centering
\vspace{+0.1in}
\textbf{Stacked Stellar Fits\phantom{AAAAAAAA}}\par
\includegraphics[width=0.48\textwidth]{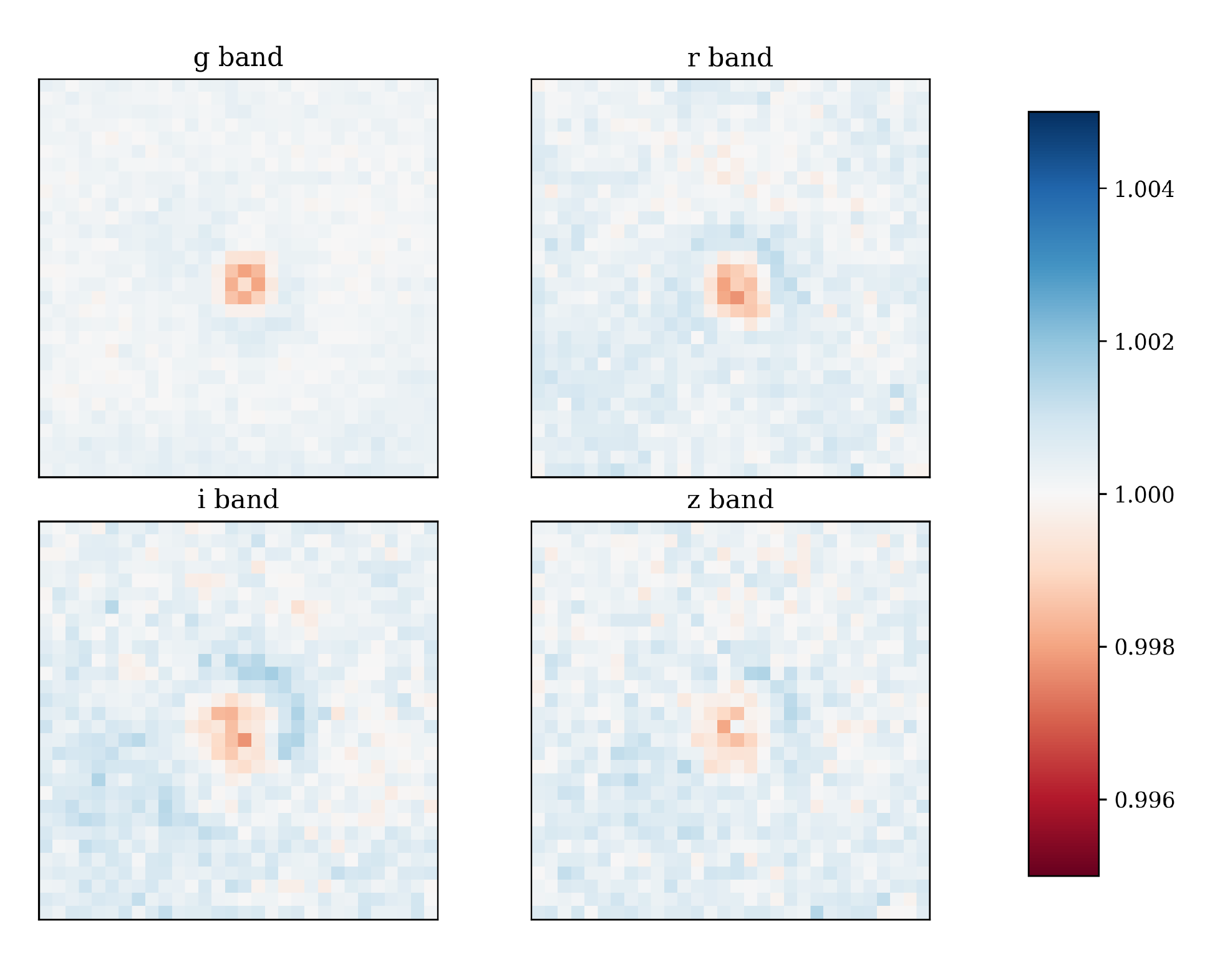}
\textbf{Stacked DES-SN Fits\phantom{AAAAAAAA}}\par
\includegraphics[width=0.48\textwidth]{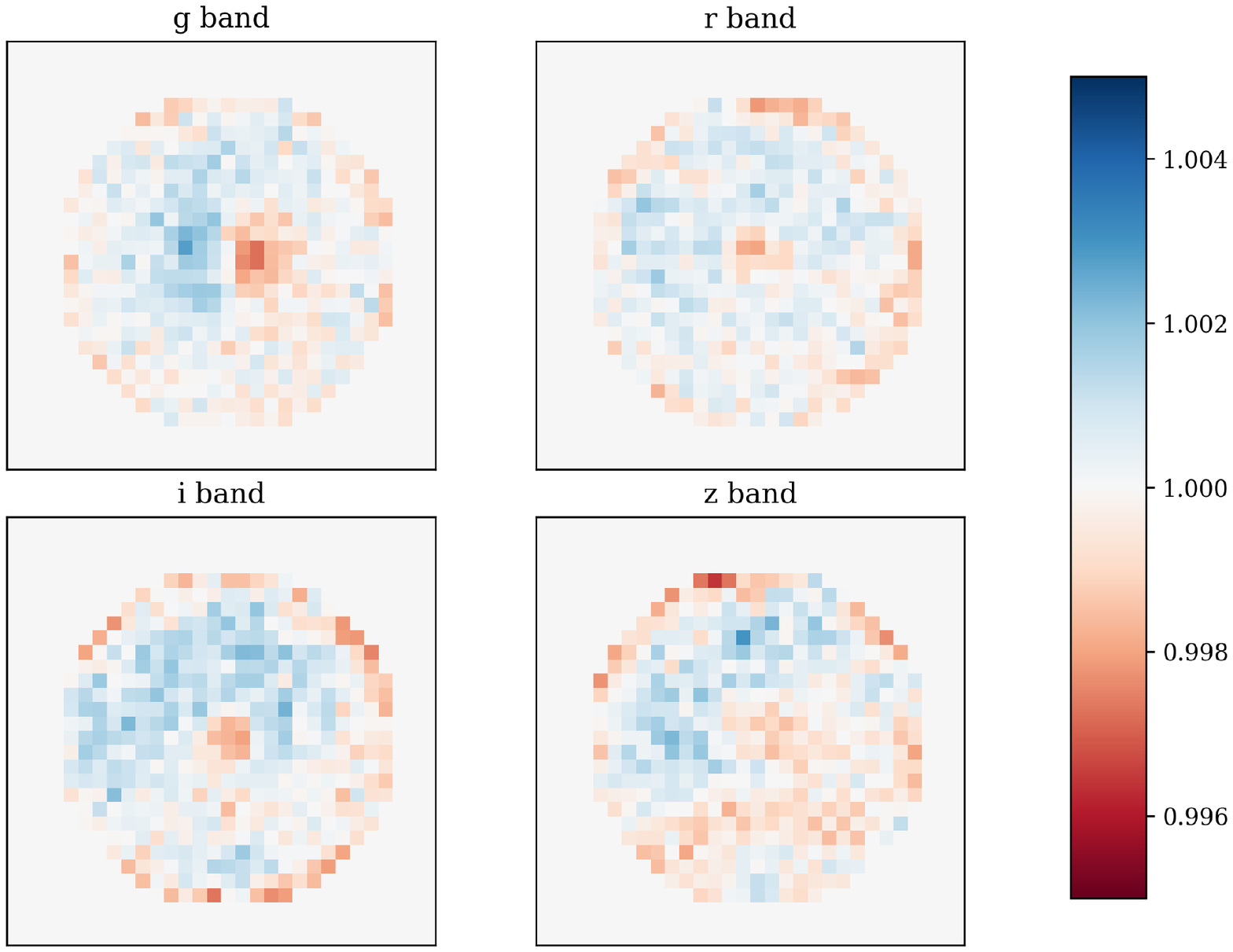}
\textbf{Stacked Fake SN Fits\phantom{AAAAAAAA}}\par
\includegraphics[width=0.48\textwidth]{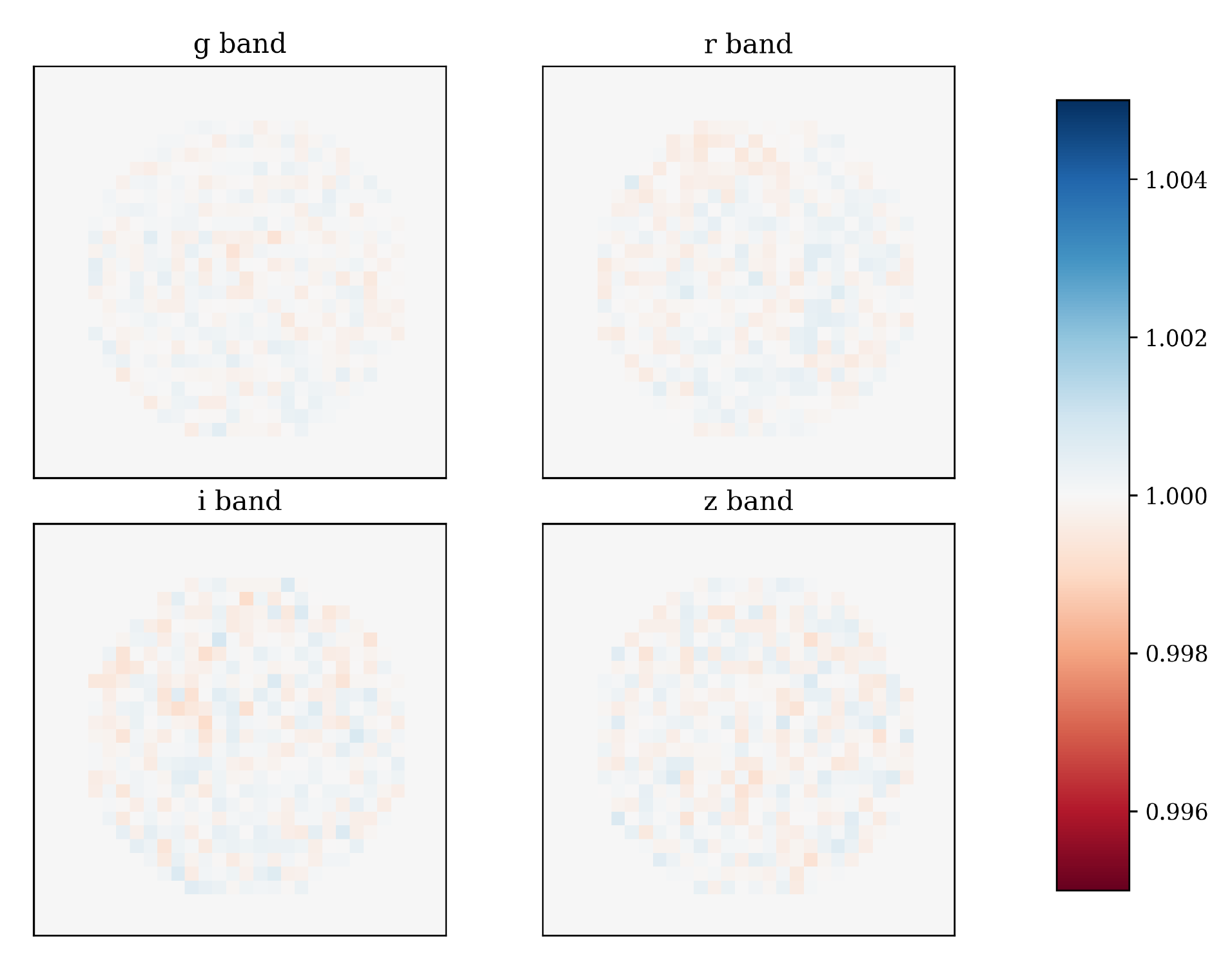}
\caption{\textbf{Top Panel:} Ratio of stellar model to DECam data image  for 3000 stacked cutouts of tertiary standard stars fainter than $19^{th}$ Mag. \textbf{Middle Panel:} Ratio of {\tt SMP} SN + galaxy model to DECam data image for 300 stacked cutouts of SNe in the DES-SN dataset brighter than $21^{st}$Mag. \textbf{Bottom Panel:} The same ratio but for the results on the fake SNe Ia. \\}
\label{percresid}
\end{figure}

\section{DES-SN Spec Sample Y1-Y3}
\label{data}

In this work, we analyze the spectroscopically confirmed SN~Ia subset of the data. As described in \cite{D'Andrea18}, 533 transients were targeted for spectroscopic classification, 251 of which were spectroscopically classified as Type Ia. We have run {\tt SMP} photometry on this sample, and show representative examples of our resulting light curves across a range of redshifts in Figure \ref{representativelc}. Light curve fits to the SALT2 model are included to guide the eye, however we refer to B18-SYS for a detailed discussion of light curve fitting and light curve quality cuts. 

A table of photometric measurements and uncertainties for the DES-SN sample is available online in machine readable format (see footnote on page 3). While all corrections to the flux uncertainties are included, we provide a separate table listing the uncertainty scales (S).

\section{Crosscheck of the PSF model}
\label{psfcrosscheck}

As discussed in Section \ref{stellarphotometry}, any differences between photometry of the standard stars and the photometry of the SNe can result in photometric biases. We explicitly check for biases in photometry due to potential inaccuracies of the measured PSF model because this is not accounted for in the analysis of the fakes. This check is performed by comparing the ratio of the stellar model stamps that were used to compute the zeropoints with the data stamps (model/data). The same model/data comparison is made for the {\tt SMP} galaxy + SN model. Any potential differences between the stellar ratios and the SN ratios could lead to biases that are not canceled out by the zeropoint. In order to obtain sufficient S/N, we stack the residuals for many fits where the SNe and stars are bright. In the top panel of Figure \ref{percresid} we stack model/data stamps for 3000 stellar fits of stars ($19 < M_{star} < 21$) over 25 nights on three different CCDs. We find that inaccuracies of the PSF model are limited to $< 0.3$\% in any given pixel. Additionally, as shown in the middle panel of Figure \ref{percresid}, we stack model/data stamps for the DES-SN SNe~Ia and their host galaxies for epochs with $19 < M_{SN} < 21$ and find similar results although it is difficult to assess given the limited statistics of the spectroscopic dataset ($\sim$300 stacked exposures). Finally, in the bottom panel of Figure \ref{percresid}, we show model/data stamps for fits to the fake SNe sample. As expected, we do not observe the same discrepancies between data and model because inaccuracies in the PSF model are not simulated in our analysis of the fakes.

To analyze the impact of the observed difference between our PSF model and the SN data, we correct the PSF model by the stacked stellar residual stamps and then re-compute stellar photometry. We find that this correction results in zeropoint differences of $< 0.5$~mmag.  Given that the small 0.5~mmag bias resulting from inaccuracies of the PSF model appear for both the tertiary standard stars and the real SNe~Ia dataset, this bias will largely cancel out and is not corrected for in this analysis.

\section{Discussion}
\label{discussion}
The {\tt SMP} pipeline developed for DES-SN models the SN host galaxy and SN transient flux simultaneously in order to extract a SN flux in each exposure. We have used 10,000 fake SN light curves overlaid onto our images to quantify potential biases in our photometry. We find that biases in photometry are limited to 3~mmag, which is small in comparison to the internal calibration uncertainties described in B18-SYS (6~mmag). Additionally, we find that errors in the PSF modeling are sub-dominant to the photometric uncertainty budget. Finally, we correct our uncertainties for the host SB dependence. 

\subsection{The Host SB Dependence}

The host SB dependence was first quantified for \texttt{DiffImg} photometry in K15 and the excess scatter is also seen in the {\tt SMP} results. Because the host SB dependence is not unique to difference imaging photometry, we conclude it does not result from the use of \texttt{SWarp} (\citealt{bertin02}) which is used to co-add exposures nor is it from \texttt{hotPants} (\citealt{hotpants}). Because the size of the dependence is similar in all bands, chromatic refraction likely plays a sub-dominant role in the host SB dependence. The source of this additional scatter is likely due to a number of confounding sources similar to the photometric repeatability floor for the stars. Atmospheric distortions contribute a chromatic increase in flux scatter and astrometric errors could introduce un-modeled uncertainty in the host galaxy itself. With improvements to the astrometric solution expected in the coming analysis of the full DES-SN 5 year dataset, we will be able to examine the dependence on astrometric quality.

\subsection{Comparison To Difference Imaging}
\begin{figure}[!h]
\begin{centering}

\includegraphics[width=0.49\textwidth]{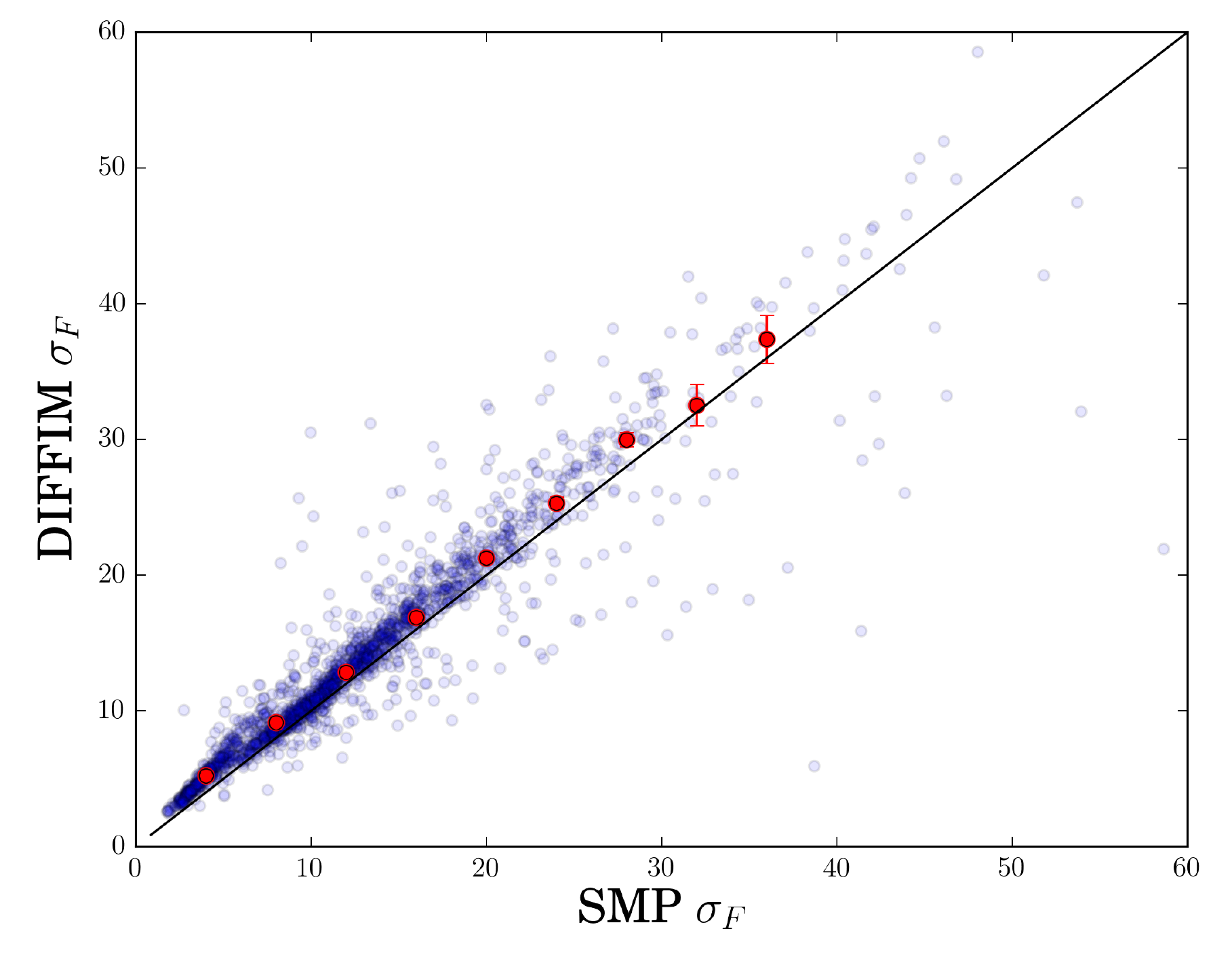}
\par\medskip\medskip
\caption{{\tt SMP} and \texttt{DiffImg} flux uncertainties with the 1-to-1 line drawn for comparison.}

\label{fig:smpvsdiffimerrs}
\end{centering}
\end{figure}

\texttt{DiffImg} was designed for DES-SN as a rapid transient identification and {\tt SMP} was designed as a precision photometric tool to be used for cosmology. Because they have been optimized for different purposes, it is difficult to make a direct comparison. We find that the fraction of catastrophic photometric outliers ($\eta_{5\sigma}$) occurs at 0.02\% for {\tt SMP} in comparison with 0.08\% for the \texttt{DiffImg} pipeline. In addition, we compare the overall size of our photometric errors and find that the uncertainties output by the {\tt SMP} pipeline are slightly smaller than those of \texttt{DiffImg} (Figure \ref{fig:smpvsdiffimerrs}).

\subsection{Future Work}

A number of improvements can be made to our photometric pipeline and analysis of the fake SNe. There are two main aspects of our fakes analysis that inhibit our ability to characterize the full extent of our photometric pipeline. First, we know the precise PSF of our fake SNe since we use the same PSF to overlay the point source and do the {\tt SMP} fitting.  In the future we will vary the PSF and calculate the impact on photometric repeatability, biases and the host SB dependence. Second, the method by which the fakes are inserted onto the images is not representative of the true astrometric uncertainty because the fakes are inserted and modeled in {\tt SMP} using the same astrometric solution. In the future we will vary the location of the fake point source on each exposure by the astrometric uncertainty. The ability to simulate both of these effects will facilitate the tracing of photometric biases due to the PSF and astrometry all the way to cosmological parameters.

For future stage IV surveys in which calibration uncertainty in the filter zeropoints approaches the $<4~$mmag level, current photometric errors (3~mmag) will need to be reduced. Additionally, as the measurement uncertainties on SN fluxes improve, it will become ever more important to understand the source of the host SB dependence. While \cite{Kessler18} show that the host SB dependence has little effect on the DES-SN detection efficiency of SNe Ia, more general transient searches for faint nearby sources (e.g. Kilonovae) on bright galaxies could also be significantly affected.

The host SB dependence may be mitigated in future DES-SN analyses with upcoming improvements to the astrometric solution and DES image processing pipelines which will include the tree ring effect noted in \cite{treerings}. 
As we do not expect the dependence to fully disappear, and to facilitate more accurate simulations of the \texttt{SMP} pipeline, we will also investigate applying a series of additive flux uncertainty floors dependent not only on $m_{SB}$, but also on observing conditions. Lastly, we will also investigate the effects of better galaxy modeling and resampling tools such as \texttt{GALSIM} (\citealt{galsim}).

\section{Conclusion} 
\label{conclusion}

We have presented the photometric pipeline for the Dark Energy Survey Supernova Program and made available the Y1-Y3 Spectroscopic SN sample light curves that are used in the cosmological analysis companion papers. This analysis uses the {\tt SMP} Pipeline to measure fluxes of SNe in their galactic environments. {\tt SMP} was run on the 251 spectroscopically confirmed SNe~Ia and was validated on a sample of 10,000 fake SNe~Ia light curves injected as point sources onto DECam images.  We find that we recover flux values to within 0.3\% accuracy. We show improvement over the \texttt{DiffImg} pipeline used for real-time transient discovery, however we find that we still must correct for the underestimated uncertainties in high local surface brightness galaxies. The {\tt SMP} pipeline will be tested further on 40,000 fake SNe and ultimately run on the full five year photometrically classified dataset of  $\sim$3000 likely SNe~Ia.

\acknowledgments
This paper has gone through internal review by the DES collaboration. DB and MS were supported by DOE grant DE-FOA-0001358 and NSF grant AST-1517742. This research used resources of the National Energy Research Scientific Computing Center (NERSC), a DOE Office of Science User Facility supported by the Office of Science of the U.S. Department of Energy under Contract No. DE-AC02-05CH11231. Part of this research was conducted by the Australian Research Council Centre of Excellence for All-sky Astrophysics (CAASTRO), through project number CE110001020.

Funding for the DES Projects has been provided by the U.S. Department of Energy, the U.S. National Science Foundation, the Ministry of Science and Education of Spain, 
the Science and Technology Facilities Council of the United Kingdom, the Higher Education Funding Council for England, the National Center for Supercomputing 
Applications at the University of Illinois at Urbana-Champaign, the Kavli Institute of Cosmological Physics at the University of Chicago, 
the Center for Cosmology and Astro-Particle Physics at the Ohio State University,
the Mitchell Institute for Fundamental Physics and Astronomy at Texas A\&M University, Financiadora de Estudos e Projetos, 
Funda{\c c}{\~a}o Carlos Chagas Filho de Amparo {\`a} Pesquisa do Estado do Rio de Janeiro, Conselho Nacional de Desenvolvimento Cient{\'i}fico e Tecnol{\'o}gico and 
the Minist{\'e}rio da Ci{\^e}ncia, Tecnologia e Inova{\c c}{\~a}o, the Deutsche Forschungsgemeinschaft and the Collaborating Institutions in the Dark Energy Survey. 

The Collaborating Institutions are Argonne National Laboratory, the University of California at Santa Cruz, the University of Cambridge, Centro de Investigaciones Energ{\'e}ticas, 
Medioambientales y Tecnol{\'o}gicas-Madrid, the University of Chicago, University College London, the DES-Brazil Consortium, the University of Edinburgh, 
the Eidgen{\"o}ssische Technische Hochschule (ETH) Z{\"u}rich, 
Fermi National Accelerator Laboratory, the University of Illinois at Urbana-Champaign, the Institut de Ci{\`e}ncies de l'Espai (IEEC/CSIC), 
the Institut de F{\'i}sica d'Altes Energies, Lawrence Berkeley National Laboratory, the Ludwig-Maximilians Universit{\"a}t M{\"u}nchen and the associated Excellence Cluster Universe, 
the University of Michigan, the National Optical Astronomy Observatory, the University of Nottingham, The Ohio State University, the University of Pennsylvania, the University of Portsmouth, 
SLAC National Accelerator Laboratory, Stanford University, the University of Sussex, Texas A\&M University, and the OzDES Membership Consortium.

Based in part on observations at Cerro Tololo Inter-American Observatory, National Optical Astronomy Observatory, which is operated by the Association of 
Universities for Research in Astronomy (AURA) under a cooperative agreement with the National Science Foundation.

The DES data management system is supported by the National Science Foundation under Grant Numbers AST-1138766 and AST-1536171.
The DES participants from Spanish institutions are partially supported by MINECO under grants AYA2015-71825, ESP2015-66861, FPA2015-68048, SEV-2016-0588, SEV-2016-0597, and MDM-2015-0509, 
some of which include ERDF funds from the European Union. IFAE is partially funded by the CERCA program of the Generalitat de Catalunya.
Research leading to these results has received funding from the European Research
Council under the European Union's Seventh Framework Program (FP7/2007-2013) including ERC grant agreements 240672, 291329, and 306478.
We  acknowledge support from the Australian Research Council Centre of Excellence for All-sky Astrophysics (CAASTRO), through project number CE110001020, and the Brazilian Instituto Nacional de Ci\^encia
e Tecnologia (INCT) e-Universe (CNPq grant 465376/2014-2).

This manuscript has been authored by Fermi Research Alliance, LLC under Contract No. DE-AC02-07CH11359 with the U.S. Department of Energy, Office of Science, Office of High Energy Physics. The United States Government retains and the publisher, by accepting the article for publication, acknowledges that the United States Government retains a non-exclusive, paid-up, irrevocable, world-wide license to publish or reproduce the published form of this manuscript, or allow others to do so, for United States Government purposes.

The UCSC team is supported in part by NASA grant NNG17PX03C, NSF grants AST-1518052 and 1815935, the Gordon \& Betty Moore Foundation, the Heising-Simons Foundation, and by fellowships from the Alfred P.\ Sloan Foundation and the David and Lucile Packard Foundation to R.J.F.

\bibliographystyle{apj}
\bibliography{brout}

\renewcommand{\thefigure}{A.\arabic{figure}}

\setcounter{figure}{0}

\end{document}